\documentclass[usenatbib]{mn2e}

\usepackage{graphicx}
\usepackage{mathptmx}
\graphicspath{{./Images/}}
\begin{document}

\title[A 610-MHz survey of the Lockman Hole -- I.]{A 610-MHz survey of
  the Lockman Hole with the Giant Metrewave Radio Telescope -- I.
  Observations, data reduction and source catalogue for the central
  5~deg$^{2}$}

\author[T.\ Garn et al.]{Timothy Garn\thanks{E-mail: tsg25@cam.ac.uk},
                        David A.\ Green,
                        Julia M.\ Riley,
                        Paul Alexander\\
                        Astrophysics Group,
                        Cavendish Laboratory, 19 J.~J.~Thomson Ave.,
                        Cambridge CB3~0HE}

\date{\today}
\pubyear{2008}
\maketitle
\label{firstpage}

\begin{abstract}
We present observations of the Lockman Hole taken at 610~MHz with the
Giant Metrewave Radio Telescope (GMRT).  Twelve pointings were
observed, covering a total area of $\sim$5~deg$^{2}$ with a resolution
of 6~$\times$~5~arcsec$^{2}$, position angle $+45\degr$.  The majority
of the pointings have an rms noise of $\sim$60~$\mu$Jy~beam$^{-1}$
before correction for the attenuation of the GMRT primary beam.
Techniques used for data reduction and production of a mosaicked image
of the region are described, and the final mosaic is presented, along
with a catalogue of 2845 sources detected above $6\sigma$.  Radio
source counts are calculated at 610~MHz and combined with existing
1.4-GHz source counts, in order to show that pure luminosity evolution
of the local radio luminosity functions for active galactic nuclei and
starburst galaxies is sufficient to account for the two source counts
simultaneously.
\end{abstract}


\section{Introduction}
The Lockman Hole \citep*{Lockman86} is the region of lowest H{\sc i}
column density in the sky, with the low infrared background
\citep[0.38~MJy~sr$^{-1}$ at 100~$\mu$m;][]{Lonsdale03} making this
region particularly well suited for deep infrared observations.  The
{\it Spitzer Space Telescope} \citep{Werner04} observed
$\sim$12~deg$^{2}$ of the region in 2004 as part of the {\it Spitzer}
Wide-area Infrared Extragalactic survey \citep[SWIRE;][]{Lonsdale03}.
Observations were centered on $10^{\rm h}45^{\rm m}00^{\rm s}$,
$+58\degr00'00''$ (J2000 coordinates, which are used throughout this
work) using the Infrared Array Camera \citep[IRAC;][]{Fazio04}
operating at 3.6, 4.5, 5.8 and 8~$\mu$m, and the Multiband Imaging
Photometer for {\it Spitzer} \citep[MIPS;][]{Rieke04} at 24, 70 and
160~$\mu$m.

A great deal of complementary data has been taken on the Lockman Hole
at other wavelengths in order to exploit the availability of sensitive
infrared observations.  There have been deep optical observations in
the $U$, $g'$, $r'$ and $i'$~bands taken with the Mosaic-I camera at
Kitt Peak National Observatory (KPNO), to $5\sigma$ Vega magnitude
limits of 24.1, 25.1, 24.4 and 23.7 respectively \citep{Surace05}.
There is existing near-infrared data across the region from the Two
Micron All Sky Survey \citep[2MASS;][]{Beichman03}, to $J$, $H$ and
$K_{\rm s}$~band magnitudes of 17.8, 16.5 and 16.0, and a band-merged
catalogue containing 323~044 sources from the IRAC, MIPS 24~$\mu$m,
2MASS and KPNO data has been produced \citep{Surace05}.  A photometric
redshift catalogue containing 229~238 galaxies and quasars within the
Lockman Hole has been constructed from the band-merged data
\citep{RowanRobinson08}.  Data Release Six of the Sloan Digital Sky
Survey \citep[SDSS DR6;][]{AdelmanMcCarthy07} covers the whole region
in the $ugriz$ bands, with both photometric and spectroscopic
observations available to a depth of $\sim$22~mag.  Further deep
infrared observations of the Lockman Hole are underway as part of the
Deep Extragalactic Survey section of the UK Infrared Deep Sky Survey
\citep[UKIDSS;][]{Lawrence07} in the $J$, $H$ and $K$~bands, with a
target sensitivity of $J \sim 22$~mag.  There have been deep surveys
of the Lockman Hole with the Submillimetre Common-User Bolometer Array
\citep[SCUBA;][]{Holland99} at 850~$\mu$m \citep{Coppin06}, and with
the X-ray satellites {\it ROSAT} \citep{Hasinger98}, {\it XMM-Newton}
\citep{Hasinger01,Mainieri02,Brunner08} and {\it Chandra}
\citep{Polletta06}.

A variety of radio surveys cover areas within the Lockman Hole.  The
first of these was by \citet{deRuiter97}, who observed 149 1.4-GHz
sources within an area of 0.35~deg$^{2}$, using the Very Large Array
(VLA) in C-configuration, with an rms noise level of
30-55~$\mu$Jy~beam$^{-1}$.  A similar deep observation was carried out
by \citet{Ciliegi03}, who observed 63 sources at 4.9~GHz within a
0.087~deg$^{2}$ region, using the VLA in C-configuration, with an rms
noise level of 11~$\mu$Jy~beam$^{-1}$.  More recently, \citet{Biggs06}
found 506 sources within a 0.35~deg$^{2}$ area, using the VLA at
1.4~GHz operating in the A- and B-configurations, and with an rms
noise level of 4.6~$\mu$Jy~beam$^{-1}$.  The Faint Images of the Radio
Sky at Twenty-cm \citep*[FIRST;][]{Becker95} and NRAO VLA Sky Survey
\citep[NVSS;][]{Condon98} surveys both cover the entire region at
1.4~GHz, but only to relatively shallow noise levels of 150 and
450~$\mu$Jy~beam$^{-1}$ respectively.

There is a clear need for deep radio observations over a significant
area within the Lockman Hole, in order to extend the multi-wavelength
information presently available for a large number of sources.  In
this paper, we present observations of the Lockman Hole taken at
610~MHz with the Giant Metrewave Radio Telescope
\citep[GMRT;][]{Ananthakrishnan05}, covering $\sim$5~deg$^{2}$ of sky
with a resolution of $6\times5$~arcsec$^{2}$, position angle
(PA)~$+45\degr$ and centred on $10^{\rm h}45^{\rm m}00^{\rm s}$,
$+58\degr00'00''$ to match the SWIRE coverage.  This is the third in a
series of 610-MHz GMRT surveys targeting the {\it Spitzer} deep legacy
survey regions, following the {\it Spitzer} extragalactic First Look
Survey \citep[xFLS;][]{Garn07} and the European Large Area {\it ISO}
Survey-North 1 (ELAIS-N1) survey \citep[EN1;][]{Garn08}.  This work,
in combination with the deep optical and infrared data, will be used
to study the infrared / radio correlation for star-forming galaxies
\citep*[e.g.][]{Helou85,Condon92,Garrett02,
Appleton04,Murphy06,Vlahakis07,Ibar08,Garn08IR}, and the link between
star formation and Active Galactic Nuclei (AGN) activity
\citep[e.g.][]{Richards07,Bundy07}, as well as the properties of the
faint radio source population at 610~MHz
\citep{Bondi07,Moss07,Tasse07,Garn08}.

In Section~\ref{sec:observations} we describe the observations and
data reduction techniques used in the creation of the survey.
Section~\ref{sec:results} presents the mosaic and a source catalogue
containing 2845 sources above $6\sigma$.  In
Section~\ref{sec:sourcecounts} we calculate the 610-MHz differential
source counts, and using local radio luminosity functions obtained
from the literature, we demonstrate that a model of pure luminosity
evolution is sufficient to account for the differential source counts
at 610~MHz and 1.4~GHz simultaneously.

Throughout this work a flat cosmology with $\Omega_{\Lambda} = 0.7$
and $H_{0} = 70$~km~s$^{-1}$~Mpc$^{-1}$ is assumed.

\section{Observations and data reduction}
\label{sec:observations}
\begin{figure}
  \begin{center}
  \includegraphics[width=.4\textwidth]{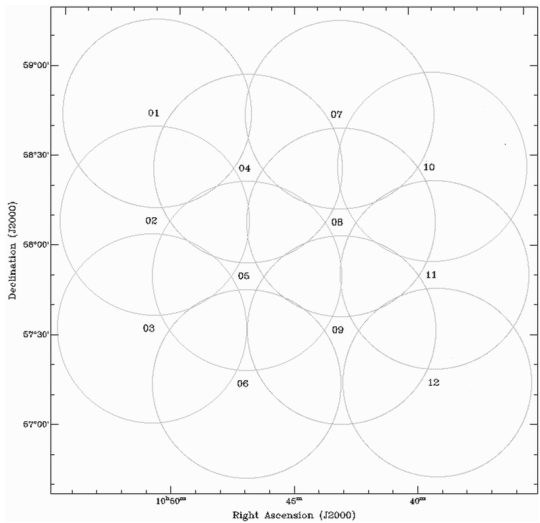}
  \caption{The arrangement of 12 pointings observed in the central
  region of the Lockman Hole.}
  \label{fig:pointings}
  \end{center}
\end{figure}

The `central' region of the Lockman Hole, consisting of 12~pointings
spaced by 36~arcmin in a hexagonal pattern (shown in
Fig.~\ref{fig:pointings}) was observed in 2004 July 24 and 25, using
the GMRT operating at 610~MHz.  A further 26 pointings surrounding the
central region (the `outer' region) were observed in 2006 July in
order to complete the coverage of the majority of the region observed
by {\it Spitzer}.  The outer observations suffered from a series of
power failures in one of the arms of the GMRT, and required several
catch-up observing sessions to cover the whole field.  This paper
describes the observations of the central 12 pointings only, with the
results from the outer region to be presented in Paper~II.

The flux density scale was set through observations of 3C48 or 3C286
at the beginning and end of each observing session.  The task {\sc
setjy} was used to calculate 610-MHz flux densities of 29.4 and
21.1~Jy respectively, using the Astronomical Image Processing Software
({\sc aips}) implementation of the \citet{Baars77} scale.  Each field
was observed for a series of interleaved scans, varying between 8 and
10~min in duration, in order to maximise the $uv$ coverage.  The scans
took place over a total time period of 7 and 9 hours on the two days,
with the typical total time being spent on each field being 48~min.  A
nearby phase calibrator, J1035$+$564, was observed for 4~min between
every three scans to monitor any time-dependent phase and amplitude
fluctuations of the telescope.  The measured phase typically varied by
less than $10\degr$ between phase calibrator observations.

\begin{figure}
  \begin{center}
    \includegraphics[width=.4\textwidth]{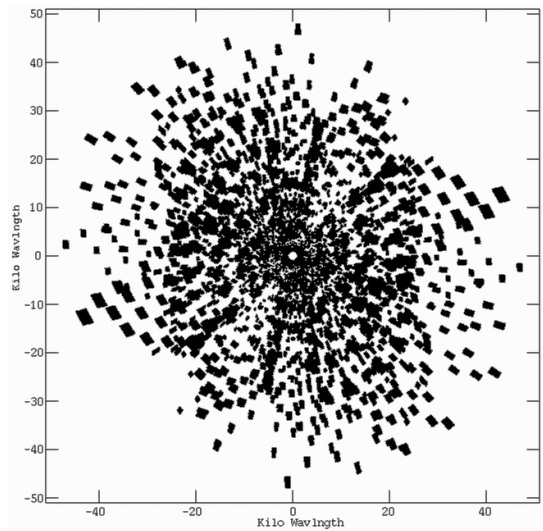}
    \caption{The $uv$ coverage for pointing 10.  Baselines less than
      1~k$\lambda$ were not used in imaging and have been omitted from
      the figure.}
    \label{fig:uvcoverage}
  \end{center}
\end{figure}

Observations were made in two 16-MHz sidebands centred on 610~MHz,
each split into 128 spectral channels, with a 16.9~s integration time.
An error in the timestamps of the {\it uv} data was corrected using
{\sc uvfxt} (\citealp[for further details see][]{Garn07}).  Initial
editing of the data was performed separately on each sideband with
standard {\sc aips} tasks, to remove bad baselines, antennas, and
channels that were suffering from large amounts of narrow band
interference, along with the first and last integration periods of
each scan.  The flux calibrators were used to create a bandpass
correction for each antenna.  In order to create a continuum channel,
five central frequency channels were combined together, and an
antenna-based phase and amplitude calibration created using
observations of J1035$+$564.  This calibration was applied to the
original data, which was then compressed into 11 channels of bandwidth
1.25~MHz, each small enough that bandwidth smearing was not a problem
for our images.  The first and last few spectral channels, which
tended to be the noisiest, were omitted from the data, leading to an
total effective bandwidth of 13.75~MHz in each sideband.  Further
flagging was performed on the 11-channel data set, and the two
sidebands combined using {\sc uvflp} \citep[again, see][]{Garn07} to
improve the $uv$ coverage.  The $uv$ coverage for one of the pointings
is shown in Fig.~\ref{fig:uvcoverage}.  Baselines shorter than
1~k$\lambda$ were omitted from the imaging, since the GMRT has a large
number of small baselines which would otherwise dominate the beam
shape.

Each pointing was divided into 31 smaller facets, arranged in a
hexagonal grid, which were imaged individually with a separate assumed
phase centre.  The large total area imaged (a diameter of $1\fdg8$,
compared with the full width at half-maximum of the GMRT, which is
$\sim$0$\fdg74$) allows bright sources well away from the observed
phase centre to be cleaned from the images, while the faceting
procedure avoids phase errors being introduced due to the non-planar
nature of the sky.  All images were made with an elliptical restoring
beam of size $6 \times 5$~arcsec$^{2}$, PA $+45\degr$, with a pixel
size of 1.5~arcsec to ensure that the beam was well oversampled.

The images went through three iterations of phase self-calibration at
10, 3 and 1 min intervals, and then a final iteration of phase and
amplitude self-calibration, at 10~min intervals, with the overall
amplitude gain held constant in order not to alter the flux density of
sources.  The self-calibration steps improved the noise level by about
10~per~cent, and significantly reduced the residual sidelobes around
the brighter sources.

\begin{figure}
  \begin{center}
  \includegraphics[width=.4\textwidth]{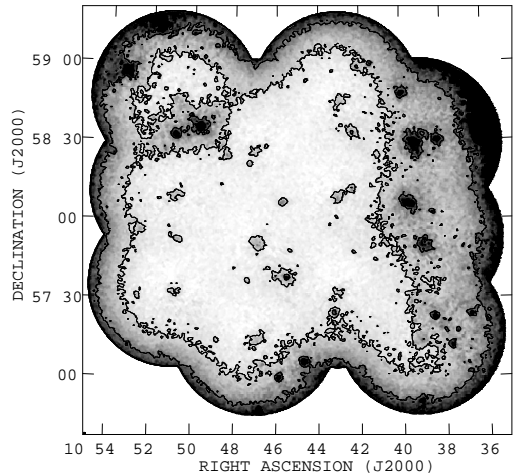}
  \caption{The rms noise of the final mosaic, calculated using {\sc
  SExtractor} (see Section~\ref{sec:results}).  The grey-scale ranges
  between 60 and 300~$\mu$Jy~beam$^{-1}$, and the contours are at 100
  and 200~$\mu$Jy~beam$^{-1}$ respectively.  The increased noise for
  pointings 10 and 11 (top right and centre-right) is due to a very
  bright radio source located at $10^{\rm h}33^{\rm m}33.2^{\rm s}$,
  $+58\degr15'02''$.}
  \label{fig:LHnoise}
  \end{center}
\end{figure}

The final rms noise before correction for the GMRT primary beam was
between 57 and 63~$\mu$Jy~beam$^{-1}$ for pointings 1 to 9 and
pointing 12, with higher noise levels of around 80~$\mu$Jy~beam$^{-1}$
for the remaining two pointings.  The expected thermal noise limit is
$\sim$55~$\mu$Jy~beam$^{-1}$, calculated using Equation~1 from
\citet{Garn07}.  The increased noise level for pointings 10 and 11 is
due to a very bright radio source (FIRST~J103333.2$+$581502), with
1.4-GHz flux density of $\sim$2~Jy, located $\sim$0$\fdg75$ from the
pointing centres.  This source lies within the region cleaned during
the imaging process, but residual sidelobes remain in our mosaic, and
the increased noise level is visible in Fig.~\ref{fig:LHnoise}.

In our previous 610-MHz GMRT surveys \citep{Garn07, Garn08} we
detected a systematic difference in the measured flux densities of
sources present in more than one pointing.  This is thought to be due
to an elevation-dependent error in the position of the GMRT primary
beam (N. Kantharia, private communication).  We repeated the analysis
of \citet{Garn07}, correcting each pointing for the nominal primary
beam shape \citep{Kantharia01} and modelling the effective phase
centre of the primary beam as having a systematic offset from its
nominal value.  A shift of $\sim$2.5~arcmin in a north-east direction
was required for each pointing in order to remove the systematic flux
density offset.

\begin{figure*}
  \begin{center}
  \includegraphics[width=\textwidth]{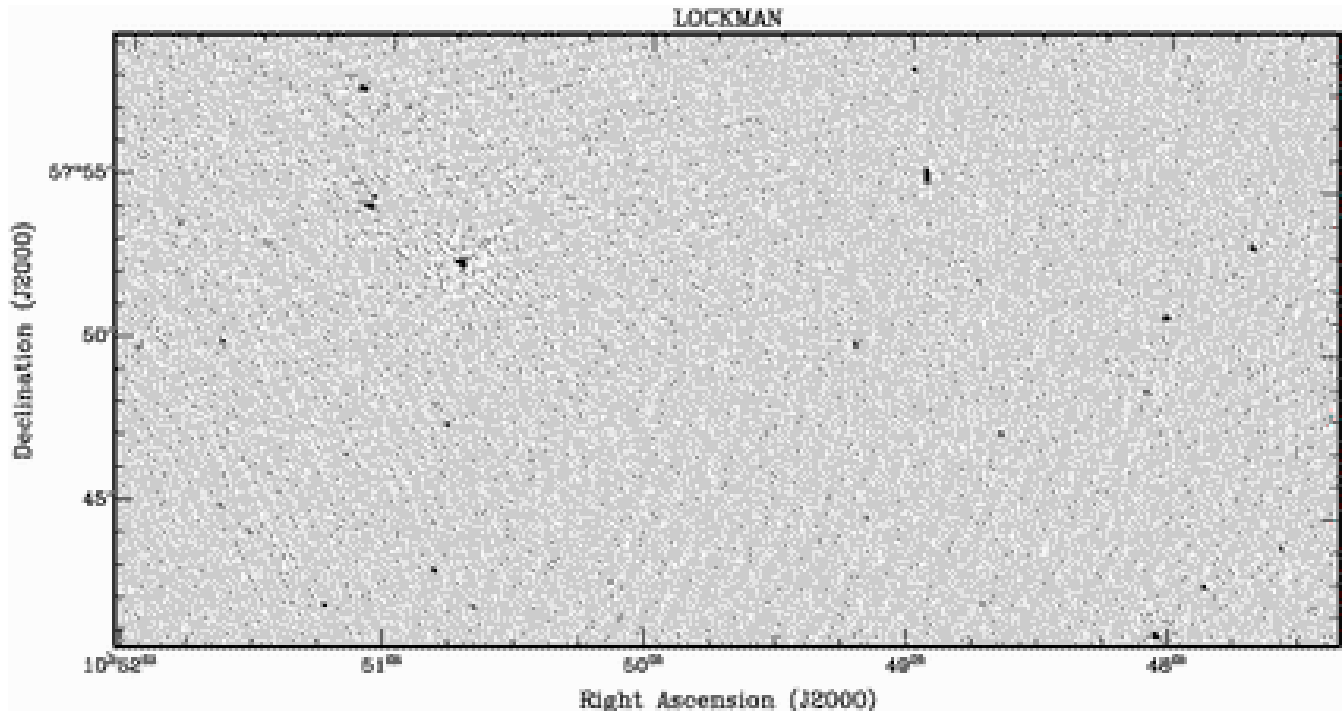}
  \caption{A 0.2~deg$^{2}$ region of the 610-MHz mosaic, illustrating
  the image quality.  The grey-scale ranges between $-0.2$ and
  1.0~mJy~beam$^{-1}$, and the noise is relatively uniform and around
  70~$\mu$Jy~beam$^{-1}$, apart from small regions near bright sources
  where the noise increases.  The brightest source in the field is
  GMRTLH J105043.2$+$575231, with a peak brightness of
  50.4~mJy~beam$^{-1}$.}
  \label{fig:LHsample}
  \end{center}
\end{figure*}

The 12 pointings were mosaicked together, taking into account the
offset primary beam and weighting the final mosaic appropriately by
the relative noise of each pointing.  The mosaic was cut off at the
point where the primary beam correction dropped to 20~per~cent of its
central value, a radius of 0\fdg53 from the centre of the outer
pointings.  The final mosaicked image is available via
http://www.mrao.cam.ac.uk/surveys/, and a sample region of the mosaic,
away from bright sources, is shown in Fig.~\ref{fig:LHsample} to
illustrate the quality of the image.

\section{Source Catalogue}
\label{sec:results}
Source Extractor \citep[{\sc SExtractor};][]{Bertin96} was
used to calculate the rms noise $\sigma$ across the mosaic.  A grid of
$16\times16$ pixels was used in order to track changes in the local
noise level, which varies significantly near the brightest sources.
Fig.~\ref{fig:LHnoise} illustrates the local noise, with the
grey-scale varying between 60~$\mu$Jy~beam$^{-1}$ (the approximate
noise level in the centre of the pointings) and
300~$\mu$Jy~beam$^{-1}$ (the expected noise level at the distance
where the GMRT primary beam gain was 20~per~cent of its central
value).  The 100 and 200~$\mu$Jy~beam$^{-1}$ contours are plotted on
Fig.~\ref{fig:LHnoise} to guide the eye.  Our GMRT data suffer from
dynamic range issues near the brightest sources, and the final mosaic
has increased noise and residual sidelobes in these regions, as for
the xFLS and EN1 surveys.  While the local noise calculated by {\sc
SExtractor} increases due to these residuals, some of them still have
an apparent signal-to-noise level that is greater than $6\sigma$.  We
therefore opted for a two-stage selection process for our final
catalogue.

\subsection{Catalogue creation}
An initial catalogue of 5223 sources was created using {\sc
SExtractor}.  The mosaic was cut off at the point where the primary
beam correction dropped to 20~per~cent of its central value, but only
sources inside the 30~per~cent region were included in the catalogue
to avoid the mosaic edges from affecting the local noise estimation.
The requirements for a source to be included in the initial catalogue
were that it had at least 5 connected pixels with brightness greater
than 3 times the local noise $\sigma$, and a peak brightness greater
than 6$\sigma$.  The image pixel size meant that the beam was
reasonably oversampled, so the source peak was taken to be the value
of the brightest pixel within a source.  The integrated flux density
was calculated using the FLUX\_AUTO option within {\sc SExtractor},
which creates an elliptical aperture around each object \citep[as
described in][]{Kron80}, and integrates the flux contained within the
ellipse.

Artefacts are seen near bright sources which may be included in the
initial catalogue.  We used the technique described in \citet{Garn08}
in order to identify regions that are affected by these artefacts, and
within these areas required sources to have a peak brightness
greater than 12$\sigma$ in order to be included in the final
catalogue.  All sources with peaks greater than 10~mJy~beam$^{-1}$
were examined for artefacts.  The final catalogue contains 2845
sources -- we have erred on the side of caution in order to produce a
catalogue with little contamination from spurious sources.

Table $\ref{tab:catalogue}$ presents a sample of 10 entries in the
catalogue, which is sorted by RA.  The full table is available as
Supplementary Material to the online version of this article, and via
http://www.mrao.cam.ac.uk/surveys/.  Column~1 gives the IAU
designation of the source, in the form GMRTLH Jhhmmss.s$+$ddmmss,
where J represents J2000.0 coordinates, hhmmss.s represents RA in
hours, minutes and truncated tenths of seconds, and ddmmss represents
the Dec.\ in degrees, arcminutes and truncated arcseconds.  Columns~2
and 3 give the RA and Dec.\ of the source, calculated using first
moments of the relevant pixel brightnesses to give a centroid
position.  Column~4 gives the brightness of the peak pixel in each
source, in mJy~beam$^{-1}$, and column~5 gives the local rms noise in
$\mu$Jy~beam$^{-1}$.  Columns~6 and 7 give the integrated flux density
and error, calculated from the local noise level and source size.
Columns 8 and 9 give the $X$, $Y$ pixel coordinates from the mosaic
image of the source centroid.  Column 10 is the {\sc SExtractor}
deblended object flag -- 1: where a nearby bright source may be
affecting the calculated flux, 2: where a source has been deblended
into two or more components from a single initial island of flux, and
3: when both of the above criteria apply.  There are 77 sources
present in our catalogue with non-zero deblend flags, and it is
necessary to examine the images to distinguish between the case where
one extended object has been represented by two or more entries, and
where two astronomically distinct objects are present.

\begin{table*}
  \caption{A sample of 10 entries from the 610-MHz Lockman Hole
  catalogue, sorted by ascending RA.  The full version of this table
  is available as Supplementary Material to the online version of this
  article, and via http://www.mrao.cam.ac.uk/surveys/.}
  \label{tab:catalogue}
  \begin{tabular}{cccccccccc} 
\hline
Name & RA & Dec.\ & Peak & Local~Noise & Int.\ Flux Density & Error & $X$ & $Y$ & Flags\\
 & J2000.0 & J2000.0 & mJy~beam$^{-1}$ & $\mu$Jy~beam$^{-1}$ & mJy & mJy & & & \\
(1) & (2) & (3) & (4) & (5) & (6) & (7) & (8) & (9) & (10) \\
\hline
GMRTLH~J104237.0$+$583759 & 10:42:37.04 & $+$58:37:59.1 &   0.564 &   82 &   1.666 &  0.136 & 3994 & 4773 & 0\\     
GMRTLH~J104237.1$+$572634 & 10:42:37.10 & $+$57:26:34.7 &   0.539 &   88 &   0.473 &  0.072 & 4018 & 1917 & 0\\     
GMRTLH~J104237.4$+$574652 & 10:42:37.41 & $+$57:46:52.1 &   0.445 &   68 &   0.412 &  0.056 & 4010 & 2729 & 3\\     
GMRTLH~J104238.3$+$572630 & 10:42:38.36 & $+$57:26:30.7 &   0.698 &   88 &   0.586 &  0.082 & 4012 & 1914 & 0\\     
GMRTLH~J104239.2$+$585011 & 10:42:39.27 & $+$58:50:11.5 &   0.536 &   73 &   0.626 &  0.085 & 3978 & 5261 & 0\\     
GMRTLH~J104239.3$+$572752 & 10:42:39.35 & $+$57:27:52.1 &   0.545 &   78 &   0.643 &  0.081 & 4006 & 1969 & 0\\     
GMRTLH~J104240.6$+$580353 & 10:42:40.62 & $+$58:03:53.6 &   0.438 &   71 &   0.786 &  0.083 & 3987 & 3409 & 0\\     
GMRTLH~J104240.6$+$573836 & 10:42:40.65 & $+$57:38:36.5 &   0.410 &   60 &   0.331 &  0.054 & 3995 & 2398 & 0\\     
GMRTLH~J104241.4$+$580505 & 10:42:41.46 & $+$58:05:05.9 &   0.648 &   77 &   0.710 &  0.082 & 3982 & 3458 & 0\\     
GMRTLH~J104241.6$+$573259 & 10:42:41.63 & $+$57:32:59.6 &   0.547 &   78 &   0.518 &  0.078 & 3992 & 2173 & 0\\     
\hline
  \end{tabular}
\end{table*}

\subsection{Comparison with the FIRST survey}
In order to test the positional accuracy of our catalogue, we paired
it with sources in the FIRST survey \citep{Becker95}.  The whole of
our 610-MHz survey region is covered by FIRST, and 340 sources are
found with positions within 6~arcsec in the two surveys.  The
difference in source positions in the GMRT catalogue relative to the
VLA FIRST survey is approximately Gaussian, with mean offset of
0.01~arcsec in RA and $-0.1$~arcsec in Dec.  The standard deviations
of the distribution are 0.3 and 0.5~arcsec respectively.  We make no
alteration to the coordinates of sources in our survey due to the good
agreement in positions.

\begin{figure}
  \begin{center}
  \includegraphics[width=.4\textwidth]{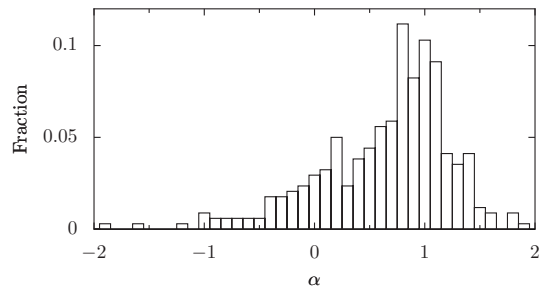}
  \caption{Radio spectral index $\alpha$ between 610~MHz and 1.4~GHz,
  for sources in the FIRST catalogue.}
  \label{fig:alpha}
  \end{center}
\end{figure}

\begin{figure}
  \begin{center}
  \includegraphics[width=.4\textwidth]{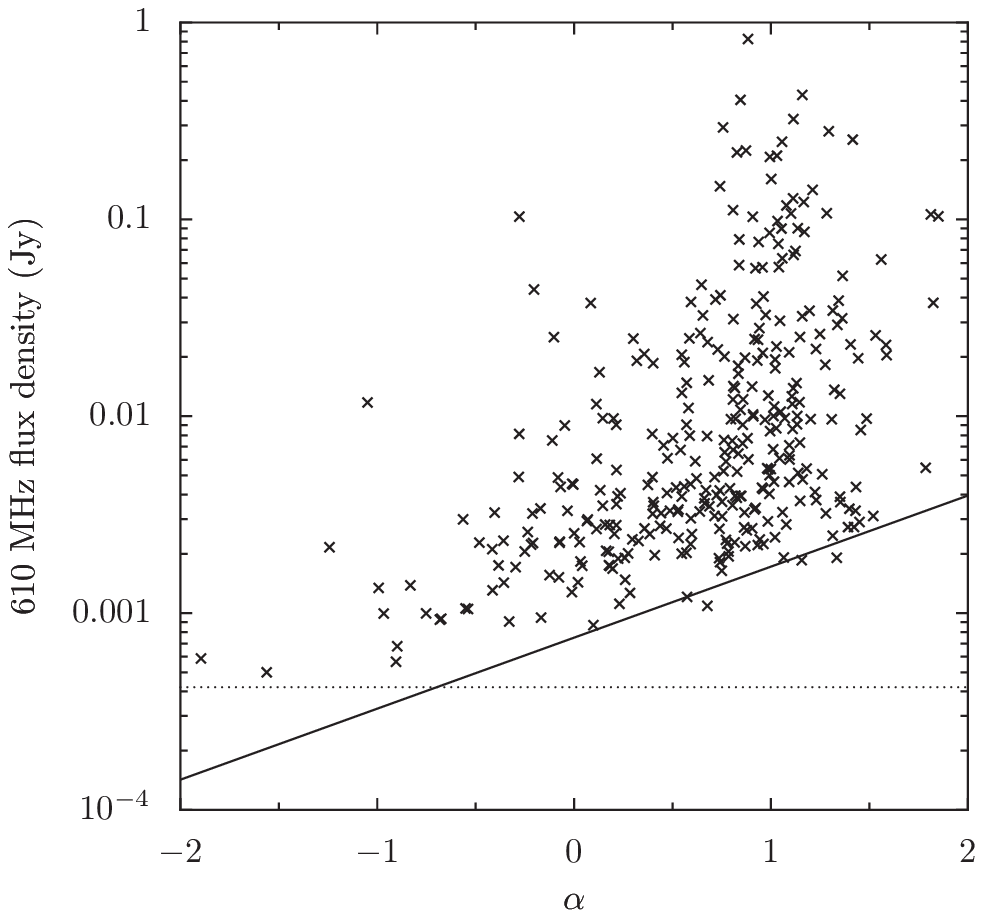}
  \caption{The variation in spectral index $\alpha$ with 610-MHz flux
  density.  Crosses represent sources in the GMRT and
  FIRST catalogues, with the solid black line showing the limiting
  spectral index that could be detected, given the respective
  sensitivity levels.  The 610-MHz flux density limit is shown by the
  dotted black line.}
  \label{fig:AlphaFlux}
  \end{center}
\end{figure}

The spectral index distribution of the matched sources is shown in
Fig.~\ref{fig:alpha}, using the integrated flux density measurements.
The distribution peaks around $\alpha=0.8$, where $\alpha$ is defined
so that the flux density $S$ scales with frequency $\nu$ as $S =
S_{0}\nu^{-\alpha}$.  There are significant biases in this
distribution due to the differing sensitivity levels of the two
surveys -- Fig.~\ref{fig:AlphaFlux} shows the variation in spectral
index with the 610-MHz flux density.  Below a few mJy, steep spectrum
sources that are visible at 610~MHz cannot be detected in the FIRST
survey, due to its limited depth (shown by the solid line).

\subsection{Sub-millimetre galaxies}
\citet{Biggs07} have produced a sample of 12 sub-millimetre galaxies
(SMGs) detected with combined VLA and Multi-Element Radio-Linked
Interferometer (MERLIN) observations at 1.4~GHz.  The majority of
sources have a flux density at 1.4~GHz less than 150~$\mu$Jy, but
there are three `bright' sources, SMG06, SMG10 and SMG11 with 1.4-GHz
flux densities of 246, 295 and 249~$\mu$Jy respectively, and typical
errors of $\sim10$~$\mu$Jy.  One of these (SMG10, at $z = 1.212$) is
detected in the 610-MHz catalogue with a flux density of
$538\pm83$~$\mu$Jy, giving it a spectral index of
$\alpha=0.72\pm0.19$.  The remaining two radio-bright SMGs, and all of
the faint ones, are undetected at 610~MHz.  The 610-MHz noise level at
the location of the two undetected bright SMGs is approximately
100~$\mu$Jy~beam$^{-1}$, leading to an upper limit on the spectral
index of these sources of $\alpha=1.05$.

\section{Differential source counts}
\label{sec:sourcecounts}
We calculated source counts for the Lockman Hole using a method
similar to the one used for the ELAIS-N1 field \citep{Garn08}, by
deriving faint and bright source counts from two separate regions of
the mosaic.  In order to calculate source counts below 10~mJy, we
chose a $75\times75$~arcmin$^{2}$ region that was relatively free from
bright sources above 10~mJy (and consequently from the errors
resulting from those sources).  For reliability, source counts were
constructed from all sources with a signal-to-noise ratio greater than
7, and binned by their integrated flux density, using the same flux
density bins as in \citet{Garn08}.  The lowest flux bin went down to
556~$\mu$Jy, corresponding to the approximate noise of
80~$\mu$Jy~beam$^{-1}$ within the selected region.  We calculated the
differential source count for sources with flux densities above 10~mJy
by considering the central $\sim$3.3~deg$^{2}$ of the map, excluding
the region near pointing 10 which had significantly greater noise.
Within this area, residual sidelobes are seen from several bright
sources, but inspection of the residuals found them to have a typical
flux density of below 1~mJy, which will not affect the source counts.

The differential source counts ${\rm d}N/{\rm d}S$ were corrected for
the fraction of the image over which they could be detected, taking
into account the increase in noise near the bright sources.  In order
to correct for resolution bias, we applied a correction factor of
3~per~cent for source counts below 1~mJy \citep{Moss07}.  No
correction for resolution bias was made for brighter sources.
Table~\ref{tab:dNdS} gives the source counts, mean flux density
$\langle S \rangle$ of sources in each bin, ${\rm d}N/{\rm d}S$ and
${\rm d}N/{\rm d}S$ normalised by $\langle S \rangle^{2.5}$, the value
expected from a static Euclidean universe, along with Poisson error
estimates calculated from the number of sources detected in each flux
density bin.  The source counts are consistent with our results from
the {\it Spitzer} extragalactic First Look Survey and ELAIS-N1 fields
\citep{Garn07,Garn08}, and in Table~\ref{tab:dNdSaverage} we give the
source counts found by combining data from the three fields.

\begin{table}
  \caption{610-MHz differential source counts for the Lockman Hole survey.}
  \label{tab:dNdS}
  \begin{tabular}{cccccc} 
  \hline
Flux Bin & $\langle S \rangle$ & $N$ & $N_{\rm c}$ & ${\rm d}N/{\rm
  d}S$ & ${\rm d}N/{\rm d}S~\langle S \rangle^{2.5}$\\
(mJy) & (mJy) & & & (sr$^{-1}$~Jy$^{-1}$) & (sr$^{-1}$~Jy$^{1.5}$)\\
  \hline
0.556 -- 0.798 & 0.674 & 94 & 108.8 & $9.4\pm1.0\times10^{8}$ & $11.1\pm1.2$\\ 
0.798 -- 1.145 & 0.951 & 61 &  61.3 & $3.7\pm0.5\times10^{8}$ & $10.4\pm1.3$\\ 
1.145 -- 1.643 & 1.381 & 31 &  31.0 & $1.3\pm0.2\times10^{8}$ & $9.3\pm1.7$\\
1.643 -- 2.358 & 1.992 & 27 &  27.0 & $7.9\pm1.5\times10^{7}$ & $14.0\pm2.7$\\ 
2.358 -- 3.384 & 2.909 & 26 &  26.0 & $5.3\pm1.0\times10^{7}$ & $24.3\pm4.8$\\
3.384 -- 4.856 & 3.981 & 14 &  14.0 & $2.0\pm0.5\times10^{7}$ & $20.0\pm5.3$\\
4.856 -- 6.968 & 5.700 &  9 &   9.0 & $8.9\pm3.0\times10^{6}$ & $21.9\pm7.3$\\ 
6.968 -- 10.00 & 8.429 &  6 &   6.0 & $4.2\pm1.7\times10^{6}$ & $27.1\pm11.1$\\
10.00 -- 13.49 & 11.70 & 14 &  14.0 & $4.0\pm1.1\times10^{6}$ & $59.3\pm15.9$\\
13.49 -- 18.20 & 15.70 &  8 &   8.0 & $1.7\pm0.6\times10^{6}$ & $52.5\pm18.6$\\
18.20 -- 24.56 & 20.71 & 17 &  17.0 & $2.7\pm0.7\times10^{6}$ & $165\pm40.0$\\ 
24.56 -- 33.14 & 27.65 & 11 &  11.0 & $1.3\pm0.4\times10^{6}$ & $163\pm49.2$\\ 
33.14 -- 44.72 & 37.70 &  7 &   7.0 & $6.0\pm2.3\times10^{5}$ & $167\pm63.1$\\ 
44.72 -- 60.34 & 57.47 &  2 &   2.0 & $1.3\pm0.9\times10^{5}$ & $101\pm71.7$\\
60.34 -- 81.41 & 74.23 &  4 &   4.0 & $1.9\pm1.0\times10^{5}$ & $285\pm142$\\ 
81.41 -- 109.8 & 94.89 &  4 &   4.0 & $1.4\pm0.7\times10^{5}$ & $390\pm195$\\
109.8 -- 148.2 & 120.5 &  3 &   3.0 & $7.8\pm4.5\times10^{4}$ & $395\pm228$\\ 
148.2 -- 200.0 & 160.3 &  1 &   1.0 & $1.9\pm1.9\times10^{4}$ & $199\pm199$\\ 
  \hline
\end{tabular}
\end{table}

\begin{table}
  \caption{Average 610-MHz differential source counts for the {\it
  Spitzer} extragalactic First Look, ELAIS-N1 and Lockman Hole surveys.}
  \label{tab:dNdSaverage}
  \begin{tabular}{cccccc} 
  \hline
Flux Bin & $\langle S \rangle$ & $N$ & $N_{\rm c}$ & ${\rm d}N/{\rm
  d}S$ & ${\rm d}N/{\rm d}S~\langle S \rangle^{2.5}$\\
(mJy) & (mJy) & & & (sr$^{-1}$~Jy$^{-1}$) & (sr$^{-1}$~Jy$^{1.5}$)\\
  \hline
0.270 -- 0.387 & 0.331 & 269 & 399.6 & $3.8\pm0.2\times10^{9}$ & $ 7.6\pm0.5$\\
0.387 -- 0.556 & 0.463 & 269 & 287.2 & $1.9\pm0.1\times10^{9}$ & $ 8.8\pm0.5$\\
0.556 -- 0.798 & 0.669 & 268 & 288.3 & $8.7\pm0.5\times10^{8}$ & $10.1\pm0.6$\\
0.798 -- 1.145 & 0.958 & 172 & 172.3 & $3.6\pm0.3\times10^{8}$ & $10.3\pm0.8$\\
1.145 -- 1.643 & 1.365 & 125 & 125.0 & $1.8\pm0.2\times10^{8}$ & $12.6\pm1.1$\\
1.643 -- 2.358 & 1.951 &  84 &  84.0 & $8.6\pm0.9\times10^{7}$ & $14.5\pm1.6$\\
2.358 -- 3.384 & 2.793 &  73 &  73.0 & $5.2\pm0.6\times10^{7}$ & $21.5\pm2.5$\\
3.384 -- 4.856 & 4.060 &  53 &  53.0 & $2.6\pm0.4\times10^{7}$ & $27.7\pm3.8$\\
4.856 -- 6.968 & 5.712 &  33 &  33.0 & $1.1\pm0.2\times10^{7}$ & $28.2\pm4.9$\\
6.968 -- 10.00 & 8.414 &  28 &  28.0 & $6.8\pm1.3\times10^{6}$ & $43.9\pm8.3$\\
10.00 -- 13.49 & 11.55 &  56 &  56.0 & $5.5\pm0.7\times10^{6}$ & $78.6\pm10.5$\\
13.49 -- 18.20 & 15.77 &  38 &  38.0 & $2.8\pm0.5\times10^{6}$ & $86.1\pm14.0$\\
18.20 -- 24.56 & 21.01 &  49 &  49.0 & $2.6\pm0.4\times10^{6}$ & $168\pm24.1$\\
24.56 -- 33.14 & 28.57 &  36 &  36.0 & $1.4\pm0.2\times10^{6}$ & $198\pm33.0$\\
33.14 -- 44.72 & 38.16 &  24 &  24.0 & $7.1\pm1.4\times10^{5}$ & $202\pm41.2$\\
44.72 -- 60.34 & 54.58 &   9 &   9.0 & $2.0\pm0.7\times10^{5}$ & $137\pm45.7$\\
60.34 -- 81.41 & 68.26 &  11 &  11.0 & $1.8\pm0.5\times10^{5}$ & $217\pm65.5$\\
81.41 -- 109.8 & 94.89 &   4 &   4.0 & $4.8\pm2.4\times10^{4}$ & $133\pm66.7$\\
109.8 -- 148.2 & 121.6 &   6 &   6.0 & $5.3\pm2.2\times10^{4}$ & $276\pm113$\\
148.2 -- 200.0 & 175.1 &   4 &   4.0 & $2.6\pm1.3\times10^{4}$ & $339\pm170$\\
  \hline
\end{tabular}
\end{table}

Fig.~\ref{fig:NumDist} shows the normalised differential source counts
from our three survey fields, along with 610-MHz source counts from
two other GMRT surveys \citep{Moss07,Bondi07}, and a collection of
shallow surveys made with the Westerbork Synthesis Radio Telescope
\citep[WSRT;][]{Valentijn77,Katgert79,Valentijn80,KatMerk85}.  The
turnover in source counts is clearly visible below $\sim$2~mJy, due to
the emergence of a new population of radio sources \citep[see
e.g.][]{Simpson06}.

\subsection{Source count models}
\label{sec:sourcecountmodels}
Radio source counts have been used to understand the evolution of
radio sources for many years \citep[e.g.][]{Longair66}.  We model the
source counts as being due to three distinct populations of radio
sources, following \citet*{Seymour04}.  The normalised differential
source counts for a population with radio luminosity function (RLF)
given by $\phi_{z}(L)$ can be calculated using
\begin{equation}
  S^{2.5} \frac{{\rm d}N}{{\rm d}S} \propto S^{1.5} \int_{0}^{\infty}
  \phi_{0}(S,z) h(z) (1 + z)^{P} {\rm d}z
\end{equation}
where $\phi_{z}(L)$ has been separated into luminosity evolution
$\phi_{0}(S,z)$ and density evolution $(1 + z)^{P}$ terms, and the
change in comoving volume with redshift is given by $h(z)$ -- for more
details, see Section~4.3 of \citet{Seymour04}.

The three populations of radio sources consist of a `steep' spectrum
AGN population with assumed spectral index of $\alpha=0.8$, a `flat'
spectrum AGN population with spectral index of 0 and a `starburst'
population, also assumed to have spectral index of 0.8.  Following
\citet{Seymour04} we consider no density evolution $(P = 0)$,
and use the 1.4-GHz AGN luminosity functions and luminosity evolution
from \citet{RowanRobinson93}, which is based on earlier work at
2.7~GHz by \citet{Dunlop90}.  The AGN contribution to the radio source
counts is therefore fixed at the same values found by previous
authors, and dominates the source counts above a few mJy.

We obtain a local RLF for starburst galaxies from \citet{Mauch07},
measured from 4006 local radio sources present in the 6dF Galaxy
Survey \citep{Jones05} and NVSS surveys, classified as starburst
galaxies through their optical spectra.  Following \citet{Hopkins98}
the starburst luminosity evolution is parameterised as $(1 + z)^{Q}$,
where we assume evolution takes place between $z = 0$ and 2, based on
measurements of the change in cosmic star formation rate across this
redshift range \citep{Madau96,Hopkins06}.  The star formation rate is
assumed to remain constant for $z > 2$.  An M82-type galaxy at
$z = 2$ would have a flux density below 1~$\mu$Jy at 1.4~GHz, and
would therefore not contribute to the observed source counts, making
this assumption valid.  

\begin{figure*}
  \begin{center}
  \includegraphics[width=.8\textwidth]{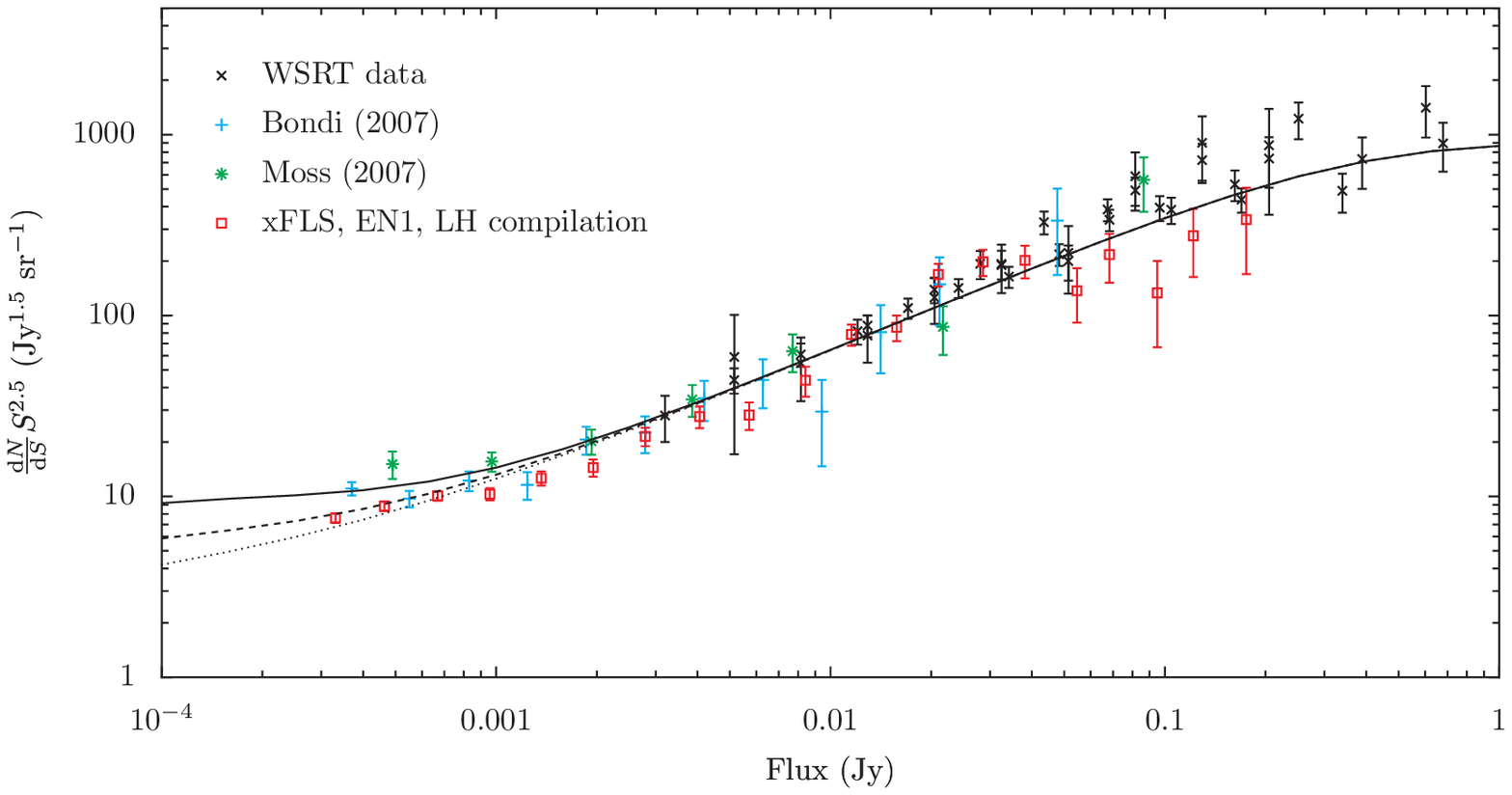}
  \caption{Differential source counts at 610~MHz
  \citep{Valentijn77,Katgert79,Valentijn80,KatMerk85,Bondi07,Moss07,Garn07,Garn08},
  normalised by the value expected in a static Euclidean universe.
  Model source counts calculated with pure luminosity evolution with
  $Q$ = 2 (dotted line), 2.5 (dashed line) and 3 (solid line) are
  shown -- see Section~\ref{sec:sourcecountmodels} for more details.}
  \label{fig:NumDist}
  \end{center}
\end{figure*}

\begin{figure*}
  \begin{center}
  \includegraphics[width=.8\textwidth]{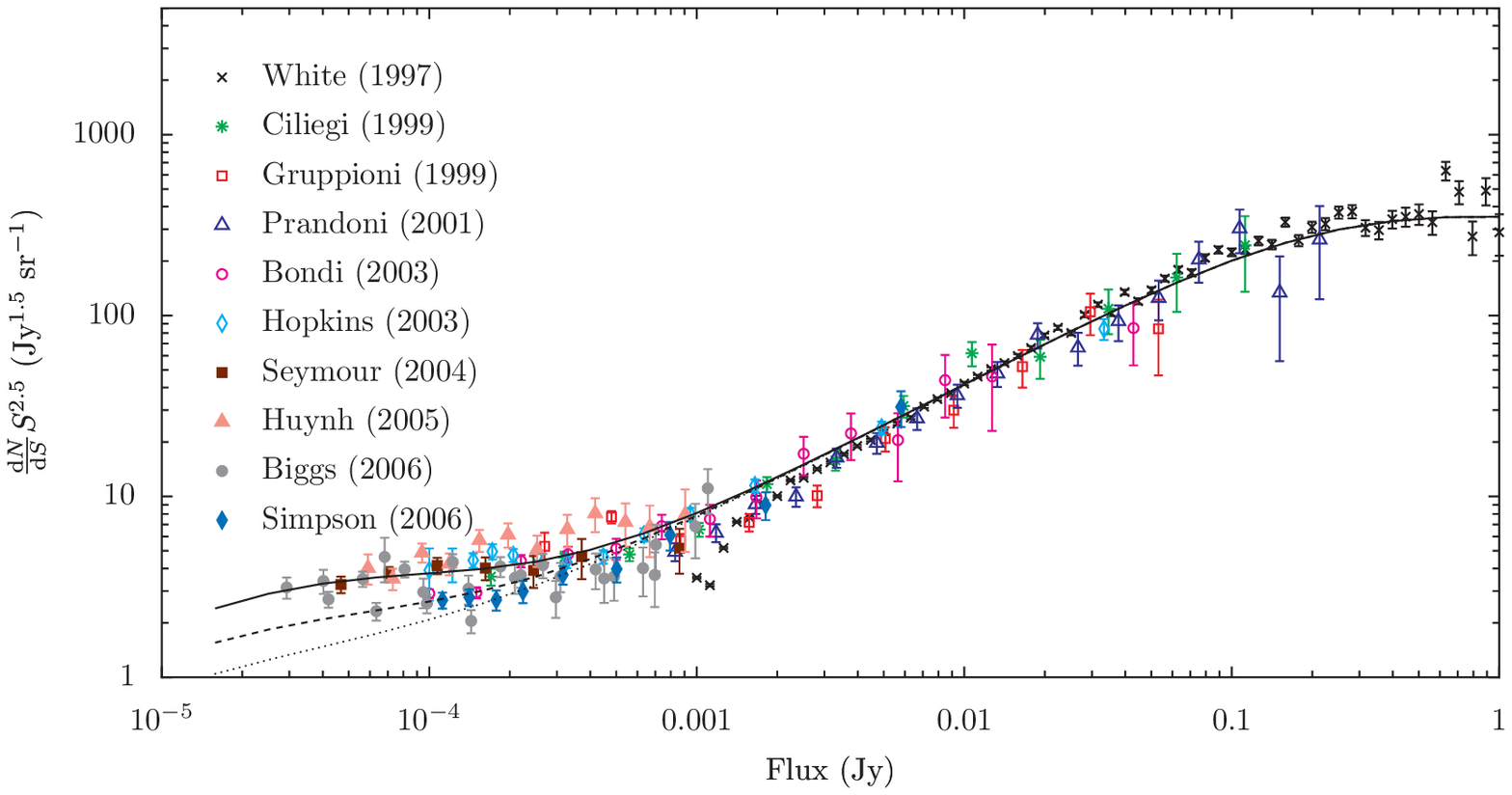}
  \caption{Differential source counts at 1.4~GHz
  \citep{White97,Ciliegi99,Gruppioni99,Prandoni01,Bondi03,Hopkins03,Seymour04,Huynh05,Biggs06,Simpson06},
  normalised by the value expected in a static Euclidean universe.
  Model source counts calculated with pure luminosity evolution with
  $Q$ = 2 (dotted line), 2.5 (dashed line) and 3 (solid line) are
  shown -- see Section~\ref{sec:sourcecountmodels} for more details.}
  \label{fig:NumDist1.4}
  \end{center}
\end{figure*}

We construct models for the source counts at 610~MHz and 1.4~GHz, with
the 1.4-GHz RLFs converted to 610~MHz through the use of the spectral
indices chosen above.  Fig.~\ref{fig:NumDist} shows the predicted
610-MHz source counts from three models of pure luminosity evolution,
with the starburst evolution parameter $Q$ taking the values 2, 2.5
and 3, in order to compare to previous works.  The source count
measurements at 1.4~GHz are taken from a wide range of surveys
\citep{White97,Ciliegi99,Gruppioni99,Prandoni01,Bondi03,Hopkins03,Seymour04,Huynh05,Biggs06,Simpson06},
and are shown in Fig.~\ref{fig:NumDist1.4} along with the same three
luminosity evolution models.

We find that a value of $Q$ between 2.5 and 3 agrees well with the
data.  From 1.4-GHz source counts, \citet{RowanRobinson93} found $Q =
3.1$, \citet*{Condon02} found $Q = 3\pm1$, \citet{Seymour04} found $Q
= 2.5\pm0.5$ and \citet{Huynh05} found $Q = 2.7$.  Using 1.4-GHz
information and a range of other star formation indicators,
\citet{Hopkins04} find $Q = 2.7\pm0.6$ and $P = 0.15\pm0.6$.
\citet{Moss07} use their 610-MHz source counts to fit a value of $Q =
2.45^{+0.25}_{-0.4}$.  Our results are therefore consistent with
previous findings, and demonstrate that both sets of source counts can
be modelled simultaneously with the same radio luminosity functions
and evolutionary dependence. 

\section{Conclusions}
We have observed $\sim$5~deg$^{2}$ of the Lockman Hole with the GMRT
at 610~MHz.  The majority of the observations have a rms noise of
$\sim$60~$\mu$Jy~beam$^{-1}$ before primary beam correction, with the
noise level increasing in the direction of a bright FIRST source
outside of our image.  A catalogue of 2845 radio source components
detected above 6$\sigma$ has been presented, and the full catalogue
and mosaic made available via http://www.mrao.cam.ac.uk/surveys/.

We calculate differential source counts at 610~MHz by considering two
separate regions of the mosaic.  The source counts are calculated
between 556~$\mu$Jy and 200~mJy, and are shown to agree with previous
610-MHz source counts in other regions.  We present average source
counts calculated from this work and two previous surveys of the {\it
Spitzer} extragalactic First Look Survey field and ELAIS-N1 field.
The turnover in differential source counts is clearly visible below
2~mJy, and we show that a three-component population containing steep
spectrum AGN, flat spectrum AGN and starburst galaxies which undergo
pure luminosity evolution is sufficient to model the 610-MHz and
1.4-GHz source counts simultaneously.

\section*{Acknowledgments}
We thank the staff of the GMRT who have made these observations
possible.  TG thanks the UK STFC for a Studentship.  The GMRT is
operated by the National Centre for Radio Astrophysics of the Tata
Institute of Fundamental Research, India.

\bibliography{./References}
\bibliographystyle{mn2e}

\label{lastpage}

\end{document}